 \documentclass[twocolumn,tighten]{aastex6}

\usepackage[T1]{fontenc}
\usepackage{amsmath}
\usepackage{color}
\usepackage{booktabs}
\usepackage{amsmath}

\newcommand\Vint{{V_{int}^{}}}
\newcommand\Lint{{L_{int}^{}}}
\newcommand\Icont{{I_{cont}^{}}}

\newcommand{\JPDF}{{JPDF}}
\newcommand{\JPDFs}{{JPDFs}}
\newcommand{\divu}{{\bf div} \, {\bf u}}

\newcommand{\Brms}{{B_{\rm rms}^{}}} 
 
 \newcommand{\taufivehundred}{{\tau_{500}^{}}}

\newcommand{\FeItwo}{{$\textrm{Fe}\,\textsc{i}\ 630.25$~nm}}

\newcommand{\perdaymmsq}{day$^{-1}$~Mm$^{-2}$}

\newcommand{\gmfe}{{the GMFE model}}
\newcommand{\BPS}{the BPS}

\newcommand{\vz}{{u_z}}

\newcommand{\ackn}{ \acknowledgements{We acknowledge support by the Spanish
    MINECO (grants AYA2011-24808, AYA2014-55078-P, CSD2007-00050,
    TIN2014-52608-REDC), NASA (NNG09FA40C, NNH15ZDA001N-HSR,
    NNX16AG90G), NSF (AST-1714955), and by the Research Council of Norway (grant
    230938/F50, Center of Excellence scheme project No.~262622, and
    computing time grants from the Programme for Supercomputing). The
    simulations have been run on clusters from the Notur project, and the
    Pleiades cluster through the computing project s1061, s1472, and s1630
    from the High End Computing (HEC) division of NASA. We also acknowledge the
    computer resources and assistance provided at the MareNostrum
    supercomputer (BSC/CNS/RES, Spain). For the 3D visualization we have used
    the VAPOR package \citep{vapor_clyne_2007}. We thank Dr.~Socas-Navarro
    for the use of the Nicole code, Drs.~Orozco-Suarez and Martinez-Gonzalez
    for clarifications on their papers, and Drs.~J.~Blanco, J.~Palacios,
    B.~Ruiz-Cobo, A.~Sainz-Dalda, and J.~Trujillo-Bueno for scientific
    discussions.}}

\shorttitle{Small-scale magnetic flux emergence in the quiet Sun}
\shortauthors{Moreno-Insertis et al.}
\begin{document}

\title{Small-scale magnetic flux emergence in the quiet Sun}

\email{Corresponding author: Fernando Moreno-Insertis, fmi@iac.es}

\author{F. Moreno-Insertis}
\affiliation{Instituto de Astrofisica de Canarias, 38205 La Laguna
  (Tenerife), Spain} 
\affiliation{Department of Astrophysics, Universidad de La Laguna, 38200 La Laguna (Tenerife), Spain}

\author{J. Martinez-Sykora}
\affiliation{Lockheed Martin Solar and Astrophysics Laboratory, Palo Alto, CA
  94304, USA} 
\affiliation{Bay Area Environmental Research Institute, NASA Research Park, Moffett Field, CA, USA}

\author{V.~H.~Hansteen}
\affiliation{Rosseland Centre for Solar Physics, University of Oslo, P.O. Box 1029 Blindern, N-0315 Oslo, Norway}
\affiliation{Lockheed Martin Solar and Astrophysics Laboratory, Palo Alto, CA
  94304, USA}

\author{D. Mu\~noz}
\affiliation{Universidad de La Laguna, 38200 La Laguna (Tenerife), Spain}

\begin{abstract}
Small bipolar magnetic features are observed to appear in the interior of
individual granules in the quiet Sun, signaling the emergence of tiny
magnetic loops from the solar interior. We study the origin of those features
as part of the magnetoconvection process in the top layers of the convection
zone. Two quiet-Sun magnetoconvection models, calculated with the
radiation-magnetohydrodynamic (MHD) Bifrost code and with domain stretching
from the top layers of the convection zone to the corona, are analyzed. Using
3D visualization as well as a posteriori spectral synthesis of Stokes
parameters, we detect the repeated emergence of small magnetic elements in
the interior of granules, as in the observations.  Additionally, we identify
the formation of organized horizontal magnetic sheets covering whole
granules. Our approach is twofold, calculating statistical properties of the
system, like joint probability density functions (\JPDFs), and pursuing
individual events via visualization tools. We conclude that the small
magnetic loops surfacing within individual granules in the observations may
originate from sites at or near the downflows in the granular and
mesogranular levels, probably in the first $1$ or $1.5$~Mm below the surface.
We also document the creation of granule-covering magnetic sheet-like
structures through the sideways expansion of a small subphotospheric magnetic
concentration picked up, and pulled out of the interior, by a nascent
granule. The sheet-like structures we found in the models may match the
recent observations of \citet{Centeno_etal_2017}.
\end{abstract}

\keywords{Sun: granulation --- Sun: photosphere ---
Sun: interior --- Convection --- Sun: magnetic fields}

\section{Introduction}\label{sec:introduction}

The advent of subarcsecond resolution in the past decade led to the
observation of concentrated magnetic flux bipoles appearing on subgranular
scales in the solar photosphere. The pioneering detections of
\citet{Centeno_etal_2007} and \citet{Martinez_Gonzalez_bellotrubio_2009} used
quiet-Sun spectropolarimetric {\it Hinode} data. The former observed the appearance
within a granule first of a patch of horizontal field of about $200$~G,
followed a few minutes later by two opposite-polarity vertical-field
patches on its edges with flux $\Phi\sim 10^{17}$~Mx. Later, the
vertical-field regions migrated toward the intergranular lanes and the
horizontal-field patch disappeared, as if a magnetic loop were surfacing
within the granule.
The latter paper observed about $70$~instances of such apparently loop-like
magnetic structures emerging inside the granules, with typical lifetime
$\lesssim 10$~minutes, $\Phi \sim 10^{16} - 2\times~10^{17}$~Mx, and
appearance rate of $1$~\perdaymmsq. Later small-scale flux emergence
observations are those by \citet{Gomory_etal_2010},
\citet{Palacios_etal_2012}, \citet{Guglielmino_etal_2012},
\citet{ortiz_etal_2014}, \citet{ortiz_etal_2016}, \citet{delacruz_etal_2015},
and \citet{Gosic_etal_2014, Gosic_etal_2016}.  Recently,
\citet{Centeno_etal_2017}, using magnetic vector measurements from the {\it
  Sunrise-II} flight, have detected the appearance of magnetic field patches
almost covering individual granules with roughly aligned, predominantly
horizontal magnetic field and footpoints near the granular edge.

Understanding the origin and time evolution of small-scale concentrated
magnetic structures requires the combination of observations and numerical
models. Within the extensive body of 3D radiation-magnetohydrodynamic (MHD) magnetoconvection
models \citep[see, e.g.][]{Nordlund_etal_2009},
only a minority of papers have focused on the phenomenon of flux emergence.
\citet{Stein_nordlund_2006} noticed magnetic fieldline bundles rising with
the granule that go out of the numerical domain through the top boundary and
leave behind a few isolated flux tubes that are coherent down to 1~Mm depth.
\citet{Cheung_etal_2007, Cheung_etal_2008, Cheung_etal_2010}
started the flux emergence process with a magnetic tube initially located
near the lower boundary or advected through it.  \citet{Cheung_etal_2007}
found that the magnetic field emerges within the interior of granules with
predominantly horizontal orientation and is expelled toward the intergranular
lanes leading to vertical-field concentrations
there. \citet{Cheung_etal_2008} provided a comparison of the numerical
results with {\it Hinode/SOT} observations, including, e.g., the formation of
transient regions with strong horizontal fields.
\citet{martinezsykora_etal_08, martinezsykora_etal_09} and
\citet{Tortosa_morenoinsertis_2009} studied the rise of the emerged
magnetized plasma to levels above the photosphere.

The actual nature, origin, and evolution of the observed subgranular features
must still be investigated. Here we report on the formation of
coherent magnetic flux structures on subgranular scales in realistic 3D
magnetoconvection experiments that can provide an explanation for the
exciting observations mentioned above. We combine direct analysis of the
numerical data, 3D visualization, and spectropolarimetric synthesis. Two
types of such structures are identified: concentrated magnetic arches
and cell-covering flux sheets.

\begin{figure*}[ht]
\hbox to \hsize{\hfill
  \includegraphics[width=\textwidth]{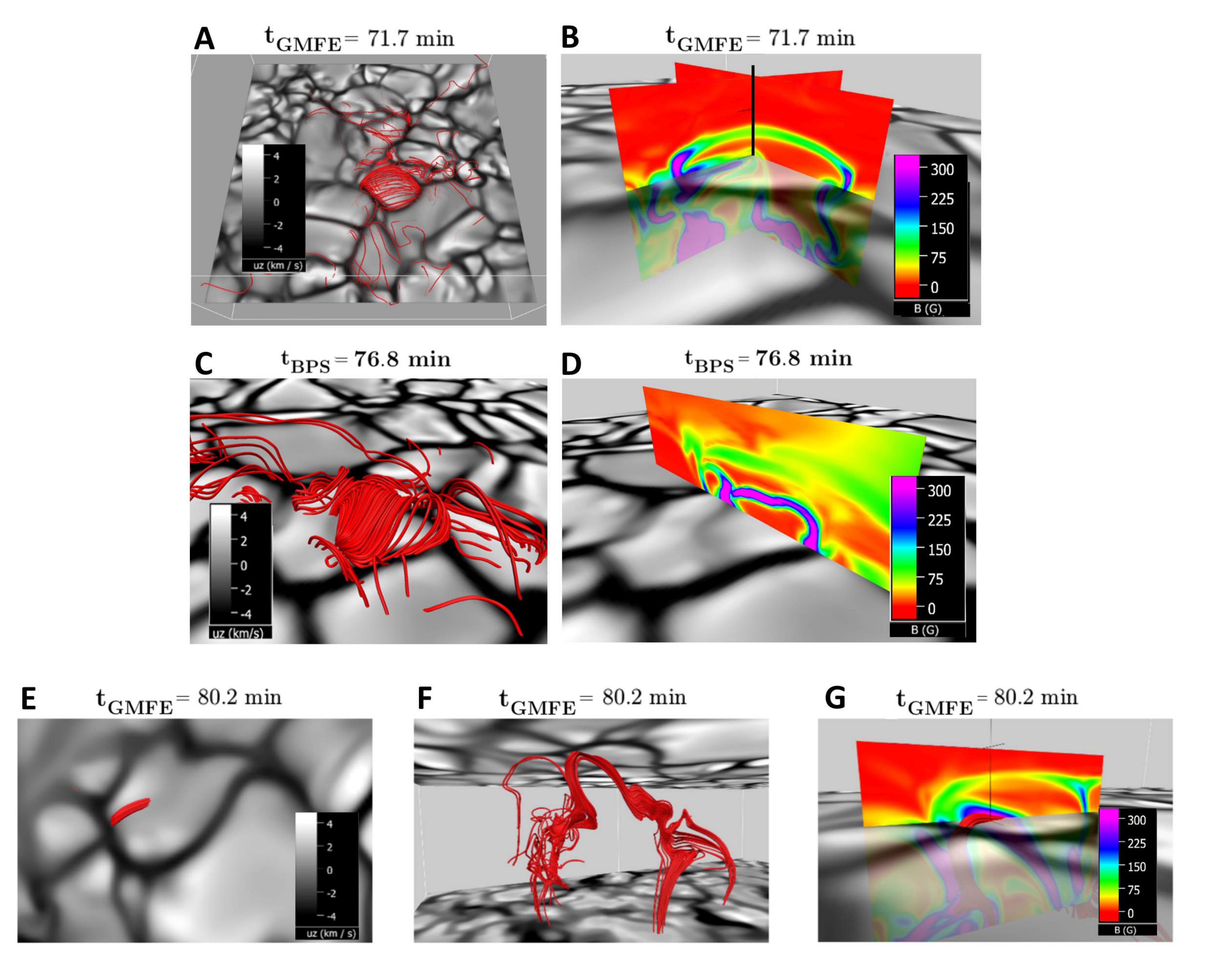}\hfill}
\caption{Panels A-D: magnetic sheet-like configurations revealed through
  fieldline tracing. A: sheet in \gmfe, with (B) two orthogonal
  field-strength maps showing vertical cross sections. (See also the
  accompanying animation). C: sheet in \BPS, with (D) an
  accompanying field strength map.  Panels E-G: emergence of an individual
  flux tube (perspective from above and below, respectively),
  with (G) corresponding $B$ map.  Grayscale maps: $\vz$ at $z=0$; panel F 
  also shows $\vz$ at $2.5$~Mm depth. The time tag indicates the simulation
  used.  (An animation of this figure is available.)}
\label{fig:fieldline_linkage}
\end{figure*}

\section{Method}\label{sec:method}

The two magnetoconvection simulations used in this Letter were calculated
with the Bifrost radiation-MHD code \citep{Gudiksen_etal_2011}.  In both, the
zero of the vertical coordinate ($z$) is set at the average height of the
$\taufivehundred=1$ corrugated surface and the bottom of the box is close to
$z=-2.5$~Mm. The first one (hereafter \gmfe) is a global magnetic flux
emergence model with a domain of $10~\times~10~\times~15$~Mm$^3$ in the
$(x,y,z)$ directions. The grid resolution is $19.5$~km horizontally and
better than $22$~km vertically for $z<2$~Mm. A simple, straight, twisted
horizontal magnetic tube with $B_{\rm axis}=7.85$~kG and total flux
$\Phi=10^{19}$~Mx is injected through the bottom boundary for $13$~solar
minutes until $t=24$~minute following the prescription of
\citet{martinezsykora_etal_08}. The convective flows drag part of the
injected field toward the surface; major arrival of flux at the surface
occurs around $t=60-70$~minutes. The events described here occur in the
subsequent $85$ minutes.  The second model is the Bifrost public simulation
of an enhanced network region (hereafter \BPS\ model) described in detail by
\citet{Carlsson_etal_2016}. Its domain is $24~\times 24~\times~16.9$~Mm$^3$;
the resolution is $48$~km horizontally and better than $19$~km vertically for
$z<5$~Mm, with snapshots available for $26.5$~solar minutes.  Instead of
injecting magnetic flux, a global bipolar potential field was implanted
throughout the box at $t=29$~minutes, allowed to be distorted by the
convection in the near-surface and subphotospheric layers, and then relax.
In the time interval studied here, both models had reached statistically
stationary values of density, temperature, and internal energy throughout the
convection zone.

Concerning the surface field, in \gmfe\ at the time of maximum vertical
unsigned flux, $\Brms=72$~G and $<B>=29$~G on the
$\taufivehundred=1$ surface.  A degraded data set (rebinned and point-spread
function (PSF)-convolved
to approximate the {\it Hinode} resolution) yields similar values. This
is within a factor $2$ of the observationally based quiet-Sun values of
\citet{Khomenko_etal_2005} or \citet{Beck_etal_2017}. On the other hand, in
\BPS, $<B>=48$ G in the photosphere.

\section{Photospheric flux emergence} \label{sec:photosphere} 

In order to discern flux emergence patterns in photospheric levels, we
calculate the magnetic linkage between photospheric elements
(Section~\ref{sec:linkage}) and Stokes polarization maps for synthetic
\FeItwo\ spectra (Section~\ref{sec:stokes}).

\begin{figure*}[ht]
\hbox to \hsize{\hfill\hskip -0mm
  \includegraphics[width=\textwidth]{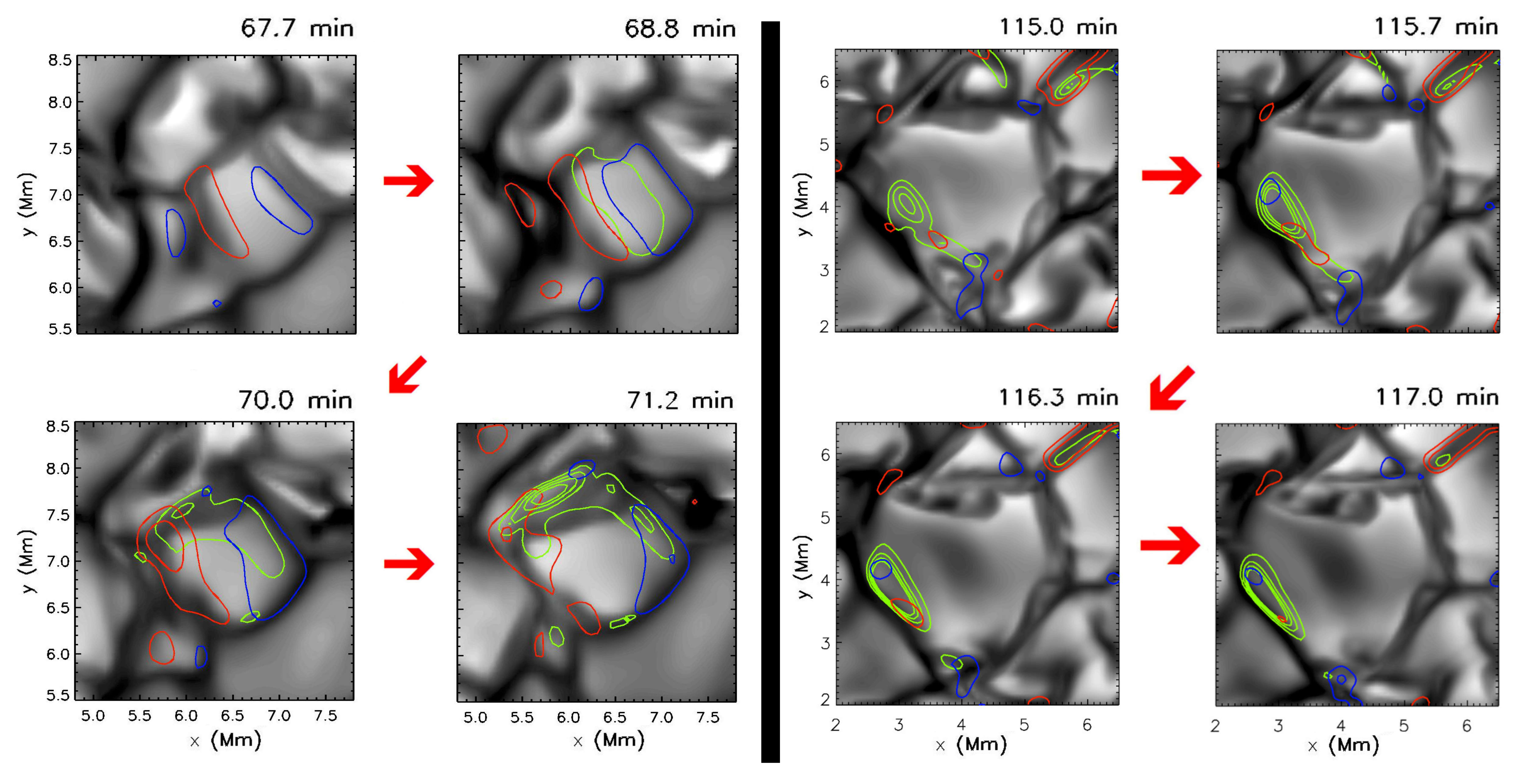}}
\caption{Time evolution of the polarization isocontours (red and blue:
  $\Vint$; green: $\Lint$) for a sheet (left) and a mini-tube (right)
  structure. Background grayscale map: the continuum intensity. Left half:
  the granule-filling sheet of Panels A-B of
  Figure~\ref{fig:fieldline_linkage}. Right half: a tube emergence event.
  (An animation of this figure is available.)}
\label{fig:sheet_tube_emergence}
\end{figure*}

\subsection{3D magnetic linkage}
\label{sec:linkage}

{\bf Sheets:} tracing magnetic field lines in the slab $0 < z < 270$~km in
\gmfe\ from random seeds biased toward strong B, we see episodes where a
granule becomes covered by an ordered magnetic-field blanket: one of them is
illustrated in Figure~\ref{fig:fieldline_linkage} (Panel A). The accompanying
animation shows the global picture. Panel B shows field strength maps on
orthogonal vertical planes that cut the blanket: the sheet is indeed a
dome-like structure of predominantly horizontal field at $z\sim 300$~km with
$B \sim 100$ G.  A sheet formation instance in \BPS\ simulation is shown on
panels C and D: field lines and (a single) field strength color map indicate
the presence of a sheet appearing in a growing granule.  Along the period
studied in \gmfe, we count at least six clear episodes of magnetic sheet
formation, i.e., of orderly blanket-like magnetic field structures
overarching a whole granule. The sheets reach basically only up to a few to
several hundred km above the photosphere, hence like standard rising plasma
elements in granules \citep[see][]{cheung_etal_inverse_granulation_2007,
  Tortosa_morenoinsertis_2009}. In most cases the magnetic sheet, after
rising a few hundred km above $z=0$, just disappears as a coherent structure
and does not outlive the granule that it covers; in at least one case, however,
the flux continues rising to the chromosphere. In \BPS\ run, we have
identified three sheet formation episodes.

\textbf{Tubes:} to try to spot emerging-tube cases inside granules via
fieldline linkage, we restrict the seeds for the tracing to emerging regions
exclusively ($\vz>0$).
An instance of emerging micro flux tube in \gmfe\ is shown in
Figure~\ref{fig:fieldline_linkage} (bottom row): viewed from above the surface
(panel E), a tube is seen emerging within a granule. Viewed from below 
(panel F), a tube-like concentration is seen down to depths of at least
several $100$~km. Using a field-strength map on a vertical plane
(panel G), we ascertain that the tube is not part of a sheet, and that it
just grazes the photosphere, where it has $B \sim 300$ G: the tube does not
rise beyond the first $100$~km above the photosphere.

The identification of rising tubes via fieldline linkage is often 
laborious. As an alternative, we next follow a
semi-observational approach using synthetic Stokes profiles.

\begin{figure*}[ht]
\hbox to \hsize{\hfill
  \includegraphics[width=\textwidth]{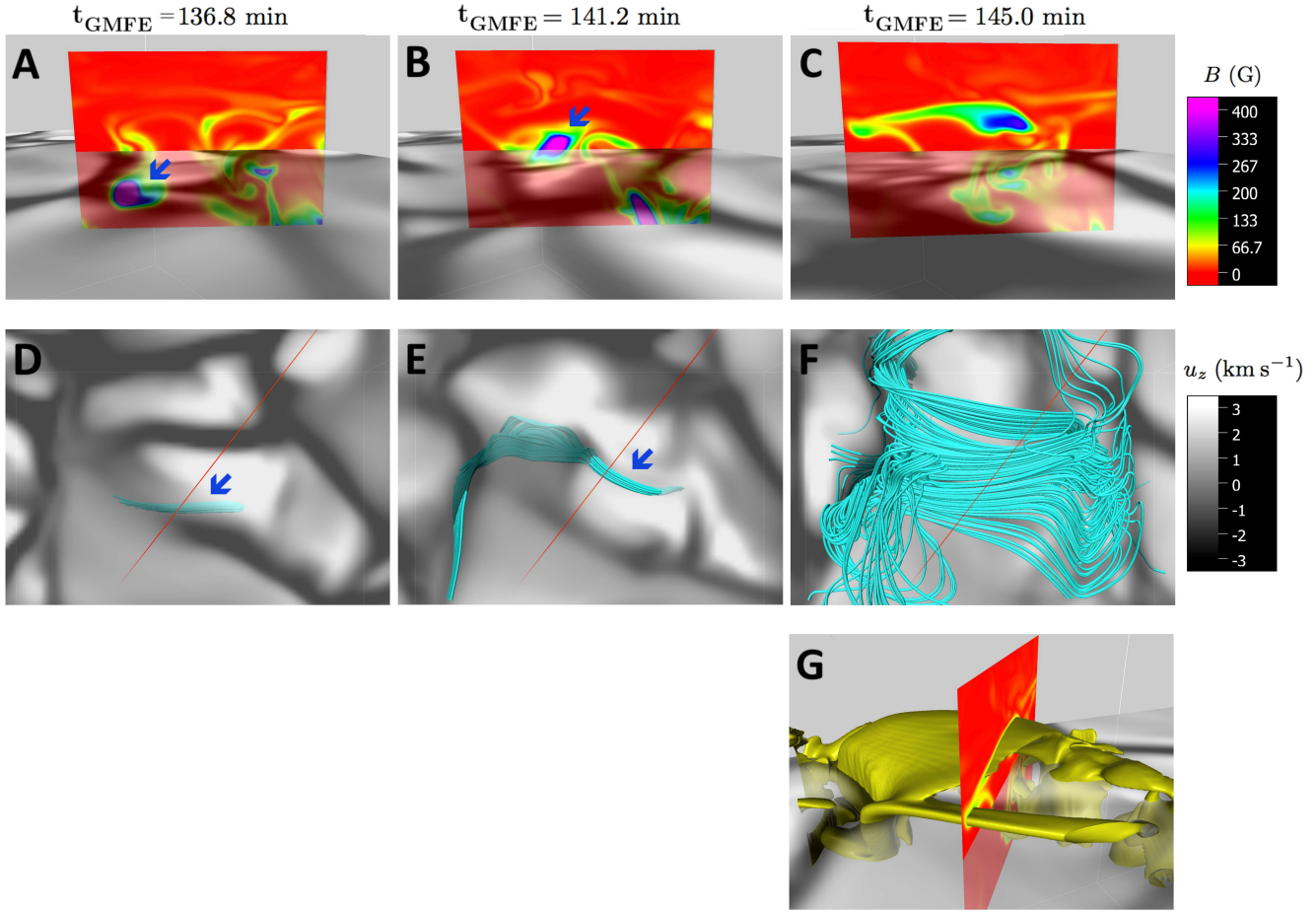}\hskip 2.55cm}

\caption{Formation of a magnetic sheet in a nascent granule. Upper row:
  vertical map for $B$.  Middle row: top view with field lines that illustrate
  the formation of the sheet.  Bottom panel: the $B=100$ G isosurface. The
  semi-transparent grayscale maps correspond to $\vz(z=0)$. The time for each
  column is given at the top.} \label{fig:sheet_formation}
\end{figure*}

\newpage
\subsection{Spectropolarimetric synthesis}\label{sec:stokes} 

For comparison with the observations, we have calculated synthetic spectra
for the \FeItwo\ line emitted by the plasma along vertical columns in
\gmfe. Using the Nicole code \citep{socasnavarro_nicolepaper_2015}, the four
Stokes parameters, $I$, $Q$, $U$, $V$, were determined assuming LTE and with
spectral resolution of $3.3$ m\AA. Following standard procedures
\citep[e.g.,][]{lites_horiz_fields_2008, sainz-dalda_2012}, for each vertical
column we define the total degree of circular and linear polarization as
\begin{eqnarray}
&\displaystyle
\Vint=\frac{1}{\lambda_r-\lambda_b}\,
\int_{\lambda_b}^{\lambda_r} \frac{|V(\lambda)|}{\Icont}\, d\lambda
\hfill \label{eq:polarization_v}\hbox to 1.7cm{\hfill}\\  
\noalign{and}
&\displaystyle
\Lint=\frac{1}{\lambda_r-\lambda_b}\,
\int_{\lambda_b}^{\lambda_r} \frac{\sqrt{Q^2(\lambda) + U^2(\lambda)}}{\Icont}\,
d\lambda\;,\hfill\label{eq:polarization_linear}
\end{eqnarray}

\noindent respectively, with $\lambda_b=630.225 $ nm, $\lambda_r=630.272$
nm, and $\Icont$ the continuum intensity near the spectral line
averaged for each snapshot.  A sign is given to $\Vint$ according to that of
the extremum of $V$ blueward of the line center.  $\Vint$ and $\Lint$ provide
a measure for vertical and transverse magnetic field strength, respectively,
in the line-forming heights ($\taufivehundred \sim 0.01$ for this line, see
\citealt{balthasar_line_formation_taus_1988, basilio_josecarlos_94}). With
$\Vint$ and $\Lint$ we construct polarization maps and try to detect the
appearance of coherent magnetic structures in the photosphere.

Figure~\ref{fig:sheet_tube_emergence} (left half) shows a time sequence of
isoline maps for $\Vint$ (red=positive, blue=negative, two contours each for
$|\Vint|=6\times 10^{-3}$ and $3\times 10^{-3}$) and $\Lint$ (green, four
contours equally spaced up to $\Lint=4\times 10^{-3}$) along $3.5$~minutes
on the background of a grayscale map of the total intensity $\Icont$ for the
magnetic sheet shown in the top left panel of
Figure~\ref{fig:fieldline_linkage}. The circular and linear polarization
signals appear basically simultaneously; the latter seem to fill the granule
shortly after first identification, as if a magnetic sheet had covered it.
In the accompanying animation, other sheet cases can be identified;
they seem to appear quasi-simultaneously with the granule with a
surface-filling nature.

Additionally, there are many instances in which the Stokes contours seem to
indicate that a coherent, tiny flux loop is emerging through the
photosphere. A clear case from \gmfe\ is illustrated in
Figure~\ref{fig:sheet_tube_emergence} (right half): a linear polarization
pattern appears in the granule interior, which $40$~s later is flanked
by two circular polarization patches (top panels).  Those patches split apart
in time, the structure reaches the intergranular lane (bottom panels), and
about a minute later it disappears.

We have found at least $19$~tube-like subgranular emergence events in the
Stokes maps in \gmfe\ (see animation for Figure~2).  Although suggestive of
an emerging loop or sheet, the indications gained through the Stokes
polarization maps should be confirmed through 3D visualization including
deeper levels, as done below.

\newpage
\section{Subphotospheric evolution}\label{sec:subphotosphere}

\begin{figure*}[ht]
\includegraphics[width=0.95\textwidth]{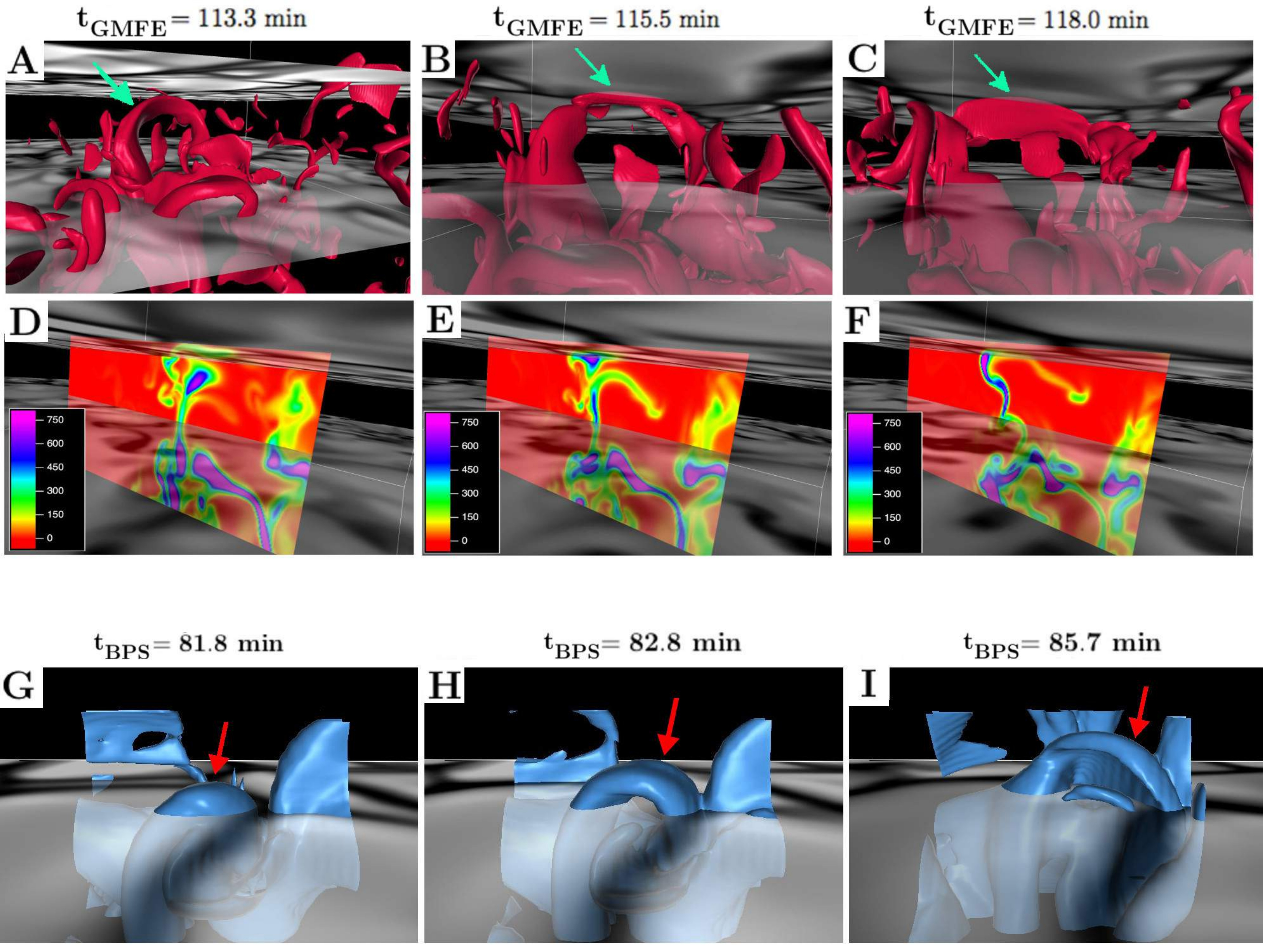}
\caption{Panels A-F: time evolution of an arch-like field concentration that
  produces the Stokes contours of Figure~\ref{fig:sheet_tube_emergence} (right
  half). Top row: $B=500$-G isosurfaces. 
The two grayscale maps correspond to $\vz$ at, respectively, $z=0$ (solid)
and $z=-1.5$~Mm (semi-transparent).
Middle row: $B$-maps (in Gauss) on vertical planes cutting across the
magnetic arches. 
Panels G-I: a subgranular flux-tube emergence event from \BPS.}
\label{fig:red_magn_concentr_landscape}
\end{figure*}

\begin{figure}[ht]
\includegraphics[width=0.5\textwidth]{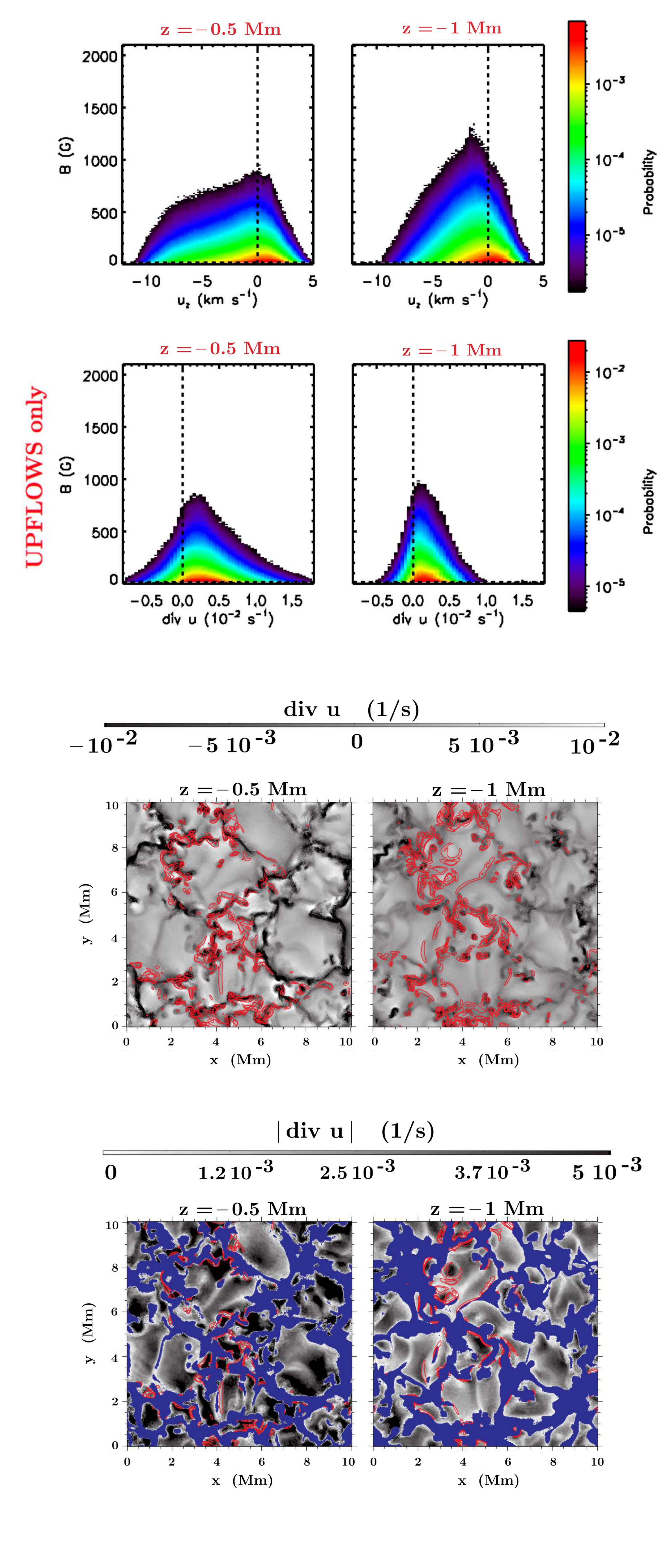}
\caption{\JPDFs\ of $B$ vs.~$\vz$ (top row), and of $B$ vs.~$\divu $ (second
  row, for upflowing plasma alone), calculated from \gmfe\ between $t=80$ -
  $140$~minutes with $10$~s cadence. Pixels with $B<20$~G are excluded.
  Lower half: grayscale maps of $\divu$ (3rd row) and $|\divu|$ (4th row)
  both at $t=101.3$~minutes and with $B=[140,280,560]$ G isocontours
  superimposed. A blue mask covers the downflows in the 4th~row.}
\label{fig:conv_pdf_one_column}

\end{figure}

\subsection{The formation of a magnetic sheet}\label{sec:formation_of_sheet}

The magnetic sheets may be naturally associated with the formation of a
granule in a subphotospheric magnetized region: the magnetized plasma will be
raised and expand sideways as the granule develops; a blanket of
ordered, horizontal field may result near the photosphere.

Figure~\ref{fig:sheet_formation} corresponds to three instants in the
development of such a sheet. A tube-like magnetic concentration had formed a
few hundred km below the photosphere: at some point it is lifted by a nascent
granule and leads to the formation of a sheet. The upper panel row contains a
side view with a vertical color map of $B$.  The middle row contains a top
view of the same snapshots: the plane of the $B$ map is now seen
edge-on. Field lines have been traced from seeds in small boxes containing
the relevant field concentrations in the $B$-maps. 
Initially (left column), the subphotospheric magnetic tube was rising toward
the photosphere, but it gets caught in a downdraft and is prevented from
rising to $z=0$. The magnetic concentration (blue arrow) is located at $z
\approx -0.43$~Mm and has a diameter of a few hundred km.  Some four minutes
later (central column) the tube has wandered off the downdraft and is part of
an incipient granule, which pulls it up through the surface. In the later
instants (rightmost panels), the magnetized plasma is lifted to heights a few
hundred km above the photosphere. In doing so, it is stretched sideways such
that a magnetic sheet is created: in panel~C the tube is still
recognizable on the right, but a roughly horizontal magnetic sheet with $B
\sim 100$ G at $z \sim 200$--$400$~km is seen to cover a granular-sized
region. The corresponding field lines (panel~F) constitute a sort
of blanket; that shape is also clear from the corresponding $B=100$ G
isosurface (panel G).

\subsection{The subsurface evolution of a tube-like feature}
\label{sec:red_isosurface_figure}

Is the bipolar feature appearing in a granule in
Figure~\ref{fig:sheet_tube_emergence} (right) really caused by a
magnetic tube rising below the surface?  
Figure~\ref{fig:red_magn_concentr_landscape} contains (top row) $B=500$~G
isosurfaces below the photosphere in a region that includes the domain of
Figure~\ref{fig:sheet_tube_emergence} (right half).  In panel~A, a prominent
arch-like feature (marked with an arrow) is seen right below the area shown
in Figure~\ref{fig:sheet_tube_emergence}, and is about to cause a bipolar
feature at the surface. Checking back in time, one can trace this arched
concentration to at least $13$~minutes earlier, when it is seen around
$1.5$~Mm depth, rising at the fringes of an upflowing convective
plume. Panel~B contains a blow-up of the central arch: its apex has already
reached the photosphere and is producing the bipolar feature of
Figure~\ref{fig:sheet_tube_emergence}. Panel~C shows the magnetic
concentration at the photosphere being {\it swallowed} by the intergranular
lane. The central panel row shows $B$-maps on vertical planes that cut across
the rising magnetic structure, confirming that the $B$-isosurfaces really
correspond to magnetic concentrations. The bottom row (panels G-I) contains
$B=150$~G isosurfaces from a subgranular tube-emergence event in \BPS; an
arched region is seen to emerge in a transverse direction to the granule
boundary; it is then pushed sideways toward the intergranule and finally
swallowed by the latter. Both in \gmfe\ and in \BPS, we have also
tested the flux-tube nature of the magnetic arches: field lines traced from
within the isosurfaces stay within them in the region occupied by the arch.

There are two important differences between the formation of sheets and tubes.
The tube-like features here had a concentrated, tube-like shape in deep
levels and were rising as part of a developed granule, whereas the structure
in Section~\ref{sec:formation_of_sheet} was picked up at shallow levels by an
incipient granule. The tube cases in the present section, although
stretched sideways when rising above the surface, did not lead to a
granule-covering magnetic sheet.

\subsection{The subphotospheric origin of the emerging flux
    concentrations} \label{sec:interior}

To locate the origin of the rising concentrated magnetic
structures detected in the experiments, we here use \gmfe\ to put
together different pieces of evidence.

(a) There is an important fraction of non-weakly magnetized plasma elements
located in upflows: the $B$-$\vz$ \JPDFs\ of
Figure~\ref{fig:conv_pdf_one_column} (top row) show that the peak probability
for the strong-field elements (say, $B \gtrsim 300$ G) at $1$~Mm depth are in
the downflow regions, as expected from magnetoconvection theory
\citep[e.g.]{Pietarila_Graham_etal_2010}, but, also,
that a significant fraction of the cases with $B$ up to
several hundred Gauss is located in upflows (which also applies to the
$z=-0.5$~Mm \JPDF).

(b) Many of those structures can survive for tens of minutes retaining high
values of the field strength. From the $B$-$\divu$ \JPDFs\ in the upflows
(second row), we see that the majority of strongly magnetized elements (say,
$B \gtrsim 300$ G) have very low (or negative) expansion rates, typically
$\divu \lesssim 3\times 10^{-3}$ s$^{-1}$.  If expanding isotropically at
that rate, $B$ would decrease by a factor $10$ in a few tens of minutes. That
is, therefore, also the typical minimum survival time that we expect for the
strong magnetic fields in those levels.

(c) The strongly magnetized volumes found in upflows are generally located in
the neighborhood of downflows. The third and fourth panel rows contain
grayscale maps of $\divu$ at $z=0.5$ and $1$~Mm with isocontours for $B=140,
280,$ and $560$~G superimposed. We see that the strong fields alineated with
the downflowing regions (third row); however, blocking the downflows with a
blue mask (fourth row), we see that part of the strong fields are moving
upwards.

We conclude that most of the stronger field concentrations are created in or
near the downflowing regions.  With lifetimes and velocities as deduced from
the \JPDFs, we estimate that the magnetic arches can go across, say, $1$~Mm.
Because additionally the upflow domains expel the plasma toward their
boundaries well before reaching the surface \citep[see][]{Stein_2012}, we see
that a concentrated magnetic structure is likely to remain in the
neighborhood of the downflow where it was created for its entire life. The
majority of flux tube examples studied in our simulations do not live longer
than $30$~minutes in upflow regions continuously; those formed below
$z=-1$~Mm are unlikely to emerge all the way to the photosphere.

\section{Discussion}\label{sec:discussion}

In this Letter, we have identified two types of small-scale emerging magnetic
structures in the solar photosphere: sheets that at maximum development cover
the granular surface (or almost do), and small magnetic tubes surfacing
within the granular cell. To do this, we have combined statistical and
visualization tools to (a) discern coherent, emerging magnetic structures on
subgranular scales in 3D numerical models of quiet-Sun magnetoconvection and
(b) investigate their origin, nature, and time evolution. A first tally
yields $19$ rising-flux tube and six magnetic sheet events in \gmfe; in
\BPS\ we count at least three flux-sheet and ten flux-tube emergence
episodes. A rough frequency of appearance of these events in the models is
therefore between $1$ and $3$~\perdaymmsq\ for tube-like events, and between
$0.3$~and~$1$~\perdaymmsq\ for sheet-like ones.  The detection of these
events in both \gmfe\ and \BPS\ is significant, as the latter is not a
global flux emergence simulation.

We  conclude that granule-covering flux sheets and tiny,
subgranular emerging tubes are different categories of emerging magnetic
structures.  The magnetic sheets seem to be created when a granule is formed
in a location traversed by a small subphotospheric magnetic tube: in the
early stages, the granules have a small size and the tube may fill an
important fraction of the nascent granule's area at the surface. The
subsequent sideways stretching and vertical compression of the magnetized
plasma volume when reaching, say, a few
hundred km above the surface leads to a flattened
magnetic structure that covers most of the granule. 
Instead, the tubes appear also in later stages of development of the granules
and do not cover their whole surface. They seem to retain a distinctive flux
tube shape from deep levels of the convective cell (e.g., from $z=1.5$~Mm)
while rising in the cell's updraft. Based in part on a statistical study we
have concluded that, in the deep levels, the rising concentrated tubes are
preferentially located in the neighborhood of the downflows and have low
expansion rates: some can rise across distances of order $1$~Mm without being
weakened by expansion nor being brought back down toward deeper levels.

On the observational side, there have been clear hints for the emergence of
small arch-like tubes within granules for about 10 years now (as detailed in
the Introduction), whereas it is only recently that \citet{Centeno_etal_2017}
may have detected granule-covering flux sheets.
On our side, we have carried out a preliminary forward-modeling exercise
through the a posteriori calculation of Stokes parameters for the
\FeItwo\ line for \gmfe\ simulation (Section~\ref{sec:stokes}). We
could draw maps of the vertical and horizontal polarization levels and locate
abundant episodes that bear clear similarities to the observations.

A full report, more detailed forward-modeling, and inclusion of further
numerical simulations are left for a future publication. 

\ackn


\end{document}